\documentstyle[12pt]{article}
\newcommand{\beq}{\begin{equation}}
\newcommand{\eeq}{\end{equation}}
\newcommand{\beqa}{\begin{eqnarray}}
\newcommand{\eeqa}{\end{eqnarray}}
\newcommand{\ba}{\begin{array}}
\newcommand{\ea}{\end{array}}
\newcommand{\half}{\frac{1}{2}}
\newcommand{\pr}{\frac{\hbar^2}{2\mu}}
\newcommand{\sqpr}{\frac{\hbar}{\sqrt{2\mu}}}

\begin{document}

\begin{flushright}
Preprint CAMTP/97-1\\
February 1997\\
\end{flushright}

\vskip 0.5 truecm
\begin{center}
\large
{\bf  Supersymmetric quantum mechanics based on higher excited states II:
a few new examples of isospectral partner potentials}\\
\vspace{0.25in}
\normalsize
Marko Robnik\footnote{e--mail: robnik@uni-mb.si}
and Junxian Liu\footnote{e--mail:junxian.liu@uni-mb.si}\\ 
\vspace{0.2in}
Center for Applied Mathematics and Theoretical Physics,\\
University of Maribor, Krekova 2, SLO--2000 Maribor, Slovenia\\
\end{center}

\vspace{0.3in}

\normalsize
\noindent
{\bf Abstract.} We apply the generalized formalism and the techniques of the
supersymmetric (susy) quantum mechanics to the cases where the superpotential
is generated/defined by higher excited eigenstates (Robnik 1997, paper I). 
The generalization is technically almost straightforward but physically 
quite nontrivial since it yields an infinity of new classes of susy-partner 
potentials, whose spectra are exactly identical except for the lowest $n+1$
states, if the superpotential is defined in terms of the $(n+1)$-st
eigenfunction, with $n=0$ reserved for the ground state. First we
show that there are practically no possibilities for shape invariant
potentials based on higher excited states. Then we calculate
the isospectral partner potentials for the following 1-dim potentials
(after separation of variables where appropriate): 
(i) 3-dim (spherically symmetric) harmonic oscillator, (ii) 3-dim 
(isotropic) Kepler problem, (iii) Morse potential, (iv) P\"oschl-Teller
type I potential, and (v) the 1-dim box potential. In all cases
except in (v) we get new classes of solvable potentials. In (v)
the partner potential to the box potential is a special case of
P\"oschl-Teller type I potential.
\vspace{0.6in}

Submitted to {\bf Journal of Physics A: Mathematical and General}
\normalsize
\vspace{0.1in}
  
\newpage

\section{Introduction}

In a recent paper (Robnik 1997, paper ({\bf I})) it has been shown that
the formalism of the supersymmetric (nonrelativistic) quantum mechanics
can be applied also to the higher excited states (say, $n$-th state)
of 1-dim potentials,  generating new partner potentials isospectral 
to the original potential, except for the lowest $n+1$ states which
are simply just missing. There we gave the example of the 1-dim harmonic
oscillator, which for all $n>0$ yields new classes of rational
potentials. In this paper we present results of a straightforward
further application of this formalism to a few most important exactly 
solvable 1-dim potentials, namely (i) spherically symmetric 3-dim
harmonic oscillator, (ii) 3-dim isotropic (spherically symmetric)
Kepler potential, (iii) Morse potential, (iv) P\"oschl-Teller type
I potential, and (v) 1-dim box potential.
\\\\
Following the seminal papers of Witten (1981) and Gendenshtein (1983) 
the methods of
supersymmetric (susy)  (nonrelativistic) quantum mechanics have quickly
developed and it has been realized, that (1)  there exist partner
potentials with precisely the same energy spectra except for the
ground state ($n=0$) (whose wavefunction  $\phi(x)=\psi_0(x)$ is used to 
generate/define the superpotential   $W(x)$ - see below)
\footnote{the ground state energy $E_0^{-}$ is missing in the partner 
Hamiltonian $H^{+}$, so that its groundstate $E_{0}^{+}=E_{1}^{-}$}, 
and that (2) if they are {\em "shape invariant"}, their spectra 
and wavefunctions can be exactly and analytically solved. 
It is believed that the list of such  susy-0 shape
invariant partner potentials is now complete and finite
(L\'evai 1989, Barclay {\em et al} 1993), and therefore
quite limited in use. The research has been later further developed
also in direction of applying the WKB methods to such classes
of Hamiltonians, including the search for improved simple
quantization conditions which would be exact in case of susy
shape invariant potentials (Barclay, Khare and Sukhatme 1993,
Barclay and Maxwell 1991, Barclay 1993, Inomata, Junker and Suparmi
1993, Junker 1995, Robnik and Salasnich 1997), 
and also in direction of exploring the applicability of
the path integral techniques (Inomata and Junker 1991,1994). 
One of the nicest presentations
of susy quantum mechanics was published by Dutt, Khare and
Sukhatme (1988), henceforth referred to as DKS. We will use their notations. 
It should be mentioned at this place that
the ideas involved behind the susy property and shape invariance
were formulated first by Infeld and Hull (1951), where they were called
the "factorization method", and these authors refer further
to the related ideas in the works of Schr\"odinger (1940,1941).
\\\\
Thus we shall use the notations of DKS, employed also in (I), and
present the brief outline of the susy-m formalism in the section 2.
Then we demonstrate in section 3 that there are practically no cases
of susy-m shape invariance for $n>0$. In sections 4-8 we present
the partner potentails of (i) through (v), and in section 9 we
draw the general conclusions and discuss the results.

\section{Generalized supersymmetric formalism}

The main point of this section is to briefly review the
formalism of the susy-n quantum mechanics, which thus can be 
generalized to arbitrary higher excited eigenstates 
$\phi(x) = \psi_n(x), \quad n=0,1,2,\dots$,
used to generate the superpotential $W(x)$, namely

\beq
W(x) = - \frac{\hbar}{\sqrt{2\mu}} \frac{\phi'}{\phi},
\label{eq:susyW}
\eeq
where $\phi'(x) = d\phi/dx$, $\mu$ is the mass of the particle
moving in the $V^{-}$ potential, $2\pi \hbar$ is the Planck constant
and $n$ is the quantum number equal to the
number of nodes of the eigenfunctions $\psi_n(x)$ of the
starting potential  $V^{-}(x)$. The energy scale is adjusted so that
the $(n+1)$-st energy eigenvalue is exactly zero, $E_n^{-} =0$.
The corresponding Hamiltonian is
$H^{-} = -\frac{\hbar^2}{2\mu} \frac{d^2}{dx^2} + V^{-}(x)$, and the
 Schr\"odinger equation reads

\beq
H^{-} \psi_n^{-} = H^{-} \phi = (-\frac{\hbar^2}{2\mu} \frac{d^2}{dx^2}
+ V^{-}(x) )\phi = 0.
\label{eq:Schrodinger}
\eeq
Obviously, because $\phi'(x)\not = 0$ at the nodes $y_j$, 
the superpotential $W(x)$ will have
singularities at the nodes $y_j$, $j=1,2,\dots,n$ of $\phi$. 
However, this does not invalidate
our derivation, but it merely means, as will become clear later on, 
that the partner potential generated by $\phi$ diverges to $+\infty$
when $x\rightarrow y_j$, for any $j=1,2,\dots,n$. This implies that
the potential wells are well defined between two consecutive 
singularities and that they do not communicate with solutions in the
neighbouring wells. Thus if $n=0$ we have the common case of
usual susy potentials defined on $(-\infty,+\infty)$, 
if $n=1$ we have two separated potential
wells, each of them on a semiinfinite domain, for $n=2$ we
have one infinite potential well on a finite domain between two
nodes $y_1$ and $y_2$, and two binding potential wells on
the two semiinfinite domains  $(-\infty,y_1]$ and $[y_2,+\infty)$,
and so on. The (partner) potentials constructed in this way
are nontrivial and certainly very interesting since they
contribute to our list of solvable potentials which now
becomes truly very rich and infinite in its contents.  
\\\\
In order to make this paper selfcontained we will build up
the formalism necessary to construct the partner potentials
and to define the shape invariance, following DKS, in
order to demonstrate that the susy formalism does not break down
anywhere on its domain of definition, 
and to define the language needed to talk about
further results that we shall present in this contribution.
\\\\
First we express the starting potential $V^{-}(x)$ in terms of
the $(n+1)$-st eigenfunction  $\phi (x) = \psi_n(x)$, by
solving (\ref{eq:Schrodinger}) 

\beq
V^{-}(x) = \frac{\hbar^2}{2\mu} \frac{\phi''}{\phi},
\label{eq:V_minus}
\eeq
which is regular everywhere, because at the nodes $y_j$ the
second derivative $\phi''(x) = d^2\phi/dx^2$ also vanishes
with $\phi$. Thus the basic Hamiltonian $H^{-}$ reads

\beq
H^{-} = \frac{\hbar^2}{2\mu}(-\frac{d^2}{dx^2} + \frac{\phi''}{\phi}).
\label{eq:Hminus}
\eeq
The two important operators are:

\beq
A^{\dagger} = \frac{\hbar}{\sqrt{2\mu}} (-\frac{d}{dx} - \frac{\phi'}{\phi}),
\label{operAplus}
\eeq
and

\beq
A = \frac{\hbar}{\sqrt{2\mu}} (\frac{d}{dx} - \frac{\phi'}{\phi}),
\label{operA}
\eeq
which gives

\beq
H^{-} = A^{\dagger}A.
\label{HminusA}
\eeq
We further {\em define the partner Hamiltonian} $H^{+}$ {\em and the partner
potential}  $V^{+}$ as

\beq
H^{+} = A A^{\dagger} = -\frac{\hbar^2}{2\mu} \frac{d^2}{dx^2} +V^{+}(x),
\label{eq:HplusA}
\eeq
where

\beq
V^{+}(x) = V^{-}(x) - \frac{\hbar^2}{\mu} \frac{d}{dx} (\frac{\phi'}{\phi})
\label{eq:Vplus}
\eeq
or

\beq
V^{+}(x) = -V^{-}(x) + \frac{\hbar^2}{\mu}(\frac{\phi'}{\phi})^2.
\label{eq:Vplusminus}
\eeq
The potentials $V^{+}$ and $V^{-}$ are called {\em susy-n partner 
potentials}.
We will show that they have the same energy levels, except for
the $(n+1)$ lowest states of $V^{-}$ for which there are no
corresponding states of $V^{+}$, so that the ground state of
the latter one is $E_0^{+} = E_{n+1}^{-}$. All higher states 
have then identical energies. From equation (\ref{eq:Vplusminus})
we see explicitly that at every node $y_j$, $j=1,2,\dots,n$ of
the defining eigenstate  $\phi=\psi_n^{-}$ the partner potential(s)
will have a singularity of the type $1/(x-y_j)^2$
such that $V^{+}(x)\rightarrow +\infty$
when $x\rightarrow y_j$, so that every branch of the partner
potential will be confining up to infinity, and the solutions
in various branches do not communicate. Thus for each $n$ we shall
find $(n+1)$ (branches of the) partner potentials.
\\\\
In terms of the superpotential $W$ defined in equation (\ref{eq:susyW})
we can write

\beq
\phi(x) = \psi_n^{-} (x) = \exp(-\frac{\sqrt{2\mu}}{\hbar} \int^x W(x)dx),
\label{eq:invW}
\eeq
which is well defined in the definition domain of any of the branches
of the partner potential, and obviously $\phi$ will go to zero where
$W$ has the poles $1/(x-y_j)$, as it should happen.
\\\\
Some of the relationships can be rewritten/reformulated in terms of
the superpotential $W(x)$ now:

\beqa
A^{\dagger} & = &-\frac{\hbar}{\sqrt{2\mu}} \frac{d}{dx} + W(x),\nonumber\\
A & = & \frac{\hbar}{\sqrt{2\mu}} \frac{d}{dx} + W(x).\nonumber\\
\label{eq:AinW}
\eeqa
Further we observe,

\beq
V^{\pm}(x) = W^2(x) \pm \frac{\hbar}{\sqrt{2\mu}} W'(x), \qquad 
W'(x)=\frac{dW}{dx},
\label{eq:VinW}
\eeq
and also

\beq
V^{+} = V^{-} + \frac{2\hbar}{\sqrt{2\mu}} \frac{dW}{dx}.
\label{eq:VplustoVminus}
\eeq
The commutator of the operators $A$ and $A^{\dagger}$ is

\beq
[A,A^{\dagger}] = \frac{2\hbar}{\sqrt{2\mu}} \frac{dW}{dx}.
\label{eq:commA}
\eeq
Now we have all tools at hand to show that the susy partner
potentials $V^{-}$ and $V^{+}$ are isospectral except for the
lowest $(n+1)$ states of $V^{-}$ which have no counterpart in
$V^{+}$, so that its ground state is  $E_0^{+} = E_{n+1}^{-}$.
\\\\
The demonstration, following DKS, is very easy: First we find that
if  $\psi_m^{-}$ is an eigenfunction of $H^{-}$ with the eigenenergy
$E_m^{-}$, then $A\psi_m^{-}$  is an eigenfunction of $H^{+}$
with the same energy:

\beq
H^{+}(A\psi_m^{-}) = AA^{\dagger}A\psi_m^{-} = AH^{-}\psi_m^{-}=
AE_m^{-}\psi_m^{-} = E_m^{-} A\psi_m^{-}.      
\label{eq:isospec}
\eeq
Now we show that this applies only to the eigenstates $m$ higher
than $n$, $m=n+1,n+2, \dots$, by considering the normalization
condition, by writing the normalized state $\psi_m^{+} = 
C_mA\psi_m^{-}$, and calculating the normalizing coefficient $C_m$, 

\beq
\| \psi_m^{+} \|^2 =  C_m^2 <A\psi_m^{-}|A\psi_m^{-}> =
 C_m^2 <\psi_m^{-}|A^{+}A\psi_m^{-}>  = C_m^2 
E_m^{-}\|\psi_m^{-}\|^2.
\label{eq:normalize}
\eeq
If all $\psi_m^{-}$ are normalized (they are certainly orthogonal,
because we deal with one dimensional systems, where degeneracies are
forbidden due to the Sturm-Liouville theorem (Courant and Hilbert 1968)
and therefore all eigenstates must be orthogonal),  then

\beq
C_m = \frac{1}{\sqrt{E_m^{-}}},
\label{eq:coeffC}
\eeq
which implies that the construction succeeds only iff $E_m^{-} > 0$,
implying that  $m>n$. Thus the two Hamiltonians $H^{-}$ and $H^{+}$
defined in (\ref{eq:Hminus}) and in (\ref{eq:HplusA}) are isospectral
except for the lowest $(n+1)$ eigenstates of $H^{-}$ which have no
counterpart in $H^{+}$.
\\\\
Counting now the eigenstates of $H^{+}$ from $m=0,1,2,\dots$, where
$m=0$ is the ground state, and $m$ is the number of nodes of the
(now also normalized) eigenfunction $\psi_m^{+}$, we have

\beq
\psi_m^{+} = \frac{1}{\sqrt{E_{n+1+m}^{-}}}A\psi_{n+1+m}^{-},
\qquad E_{m}^{+} = E_{n+1+m}^{-}.
\label{eq:normalizedpsi}
\eeq
Of course it is easy to show that, conversely, 
for every eigenstate $\psi_m^{+}$
of $H^{+}$ there exists the normalized eigenstate of $H^{-}$, namely

\beq
\psi_{n+1+m}^{-} = \frac{1}{\sqrt{E_m^{+}}} A^{\dagger}\psi_m^{+},
\qquad m=0,1,2,\dots
\label{eq:psiminusofpsiplus}
\eeq
This completes our proof of isospectrality, generalized to the case
that the generating function $\phi$ of the superpotential $W$, defined in
equation (\ref{eq:susyW}), is a higher excited wavefunction, 
namely $\phi=\psi_n^{-}$, $n=0,1,2,\dots$. As we have seen, the
formalism of superpotential and of the partner potentials works
everywhere except at the singularities located at the nodal points
$y_i$ of $\phi$, where the partner potential $V^{+}$ goes to infinity
as $1/(x-y_i)^2$, thereby defining several branches of $V^{+}$ 
well defined on their disjoint domains of definition. 
\\\\
We have demonstrated that if one of the partner systems (the Hamiltonians) 
can be solved completely (by calculating the energy levels and the 
eigenfunctions), then the susy formalism enables one to solve the
partner problem completely, following equation (\ref{eq:normalizedpsi}).
One of the most important cases is of course the harmonic oscillator,
which has been discussed in (I), whilst in this paper we study five more
examples (i)-(v) announced in Introduction.

\section{Is there shape invariance for $n>0$?}

If the solutions for the two partner Hamiltonians are both unknown, 
then another approach is necessary to solve them. In case of the
standard susy formalism with $n=0$ we have the important class of
the shape invariant potentials. 
As is well known (DKS) the shape invariance of the two partner potentials
$V^{-}$ and $V^{+}$ is defined by

\beq
V^{+}(x; a_0) = V^{-}(x;a_1) + R(a_1),
\label{eq:shapeinvariance}
\eeq
where $a_0$ is a set of parameters, $a_1=f(a_0)$ and $R(a_1)$ is
independent of $x$. The procedure is now (essentially embodied in
the factorization method of Infeld and Hull (1951)) the following.
Consider a series of Hamiltonians  $H^{(s)}$, $s=0,1,2,\dots$,
where  $H^{(0)} = H^{-}$ and $H^{(1)} = H^{+}$, by definition

\beq
H^{(s)} = -\frac{\hbar^2}{2\mu} \frac{d^2}{dx^2} + V^{-}(x;a_s) 
+\sum_{k=1}^{s} R(a_k),
\label{eq:Hs}
\eeq
where 

\beq
a_s=f^s(a_0) = \underbrace{f\circ\dots\circ f}_{s} (a_0).
\label{eq:scompf}
\eeq
Now compare the spectra of $H^{(s)}$ with $H^{(s+1)}$, and 
find

\beqa
H^{(s+1)} & = & -\frac{\hbar^2}{2\mu} \frac{d^2}{dx^2}  + V^{-}(x;a_{s+1}) + 
\sum_{k=1}^{s+1}R(a_k),\nonumber\\
H^{(s+1)} & = & -\frac{\hbar^2}{2\mu} \frac{d^2}{dx^2} + V^{+}(x;a_{s}) + 
\sum_{k=1}^{s}R(a_k),
\label{eq:partnerHs}
\eeqa
Thus it is obvious that $H^{(s)}$ and $H^{(s+1)}$ are susy partner
Hamiltonians, and they have the same spectra from the first level
upwards except for the ground state of $H^{(s)}$ whose energy
is

\beq
E_0^{(s)} = \sum_{k=1}^{s} R(a_k).
\label{eq:groundofHs}
\eeq
When going back from $s$ to $(s-1)$ we reach $H^{(1)}=H^{+}$ and
$H^{(0)} = H^{-}$, whose ground state energy is zero and its $m$-th
energy level being coincident with the ground state of the Hamiltonian
$H^{(m)}$, $m=1,2,\dots$. Therefore the complete spectrum of
$H^{-}$ is 

\beq
E_m^{-} = \sum_{k=1}^{m} R(a_k), \qquad E_0^{-} = 0.
\label{eq:finalspec}
\eeq

\vspace{0.2in}

\noindent
The generalization of shape invariance to the case of any $n\ge 0$ is 
straightforward, but it results in higher complexity and therefore
it is more rarely satisfied by the specific systems. By repeating the
above argumentation we reach the conclusion that, when 
(\ref{eq:shapeinvariance}) is satisfied for a superpotential $W$
with given $n$, then we cannot calculate the entire spectrum of
the shape invariant potential/Hamiltonian $H^{-}$, 
but only the subset (subsequence) of period $n+1$, namely

\beq
E_{n+m(n+1)}^{-} = \sum_{k=1}^m R(a_k), \qquad E_n^{-}=0, m=1,2,\dots.
\label{eq:mshapeinv}
\eeq
In the special case $n=0$ we of course recover the formula
(\ref{eq:finalspec}). For $n>0$ we have {\em none example} of
susy-n shape invariance so far. In fact we shall show now that 
there is no susy-n shape invariance for $n>0$, 
unless we find some rare exceptions.
\\\\
In fact we can see in equation (\ref{eq:shapeinvariance}) that if
$V^{-}(x;a_1)$ has no singularities as a function of $x$, then
the partner potential $V^{+}(x;a_0)$ also has no singularities,
because $R(a_1)$ is just a constant and independent of $x$. Therefore,
for $n>0$ a potential $V(x)$ cannot be shape invariant if it 
has no singularities. The only possibility then is that
$V^{-}(x;a_1)$ has singularities at the same places $y_i$ as
the partner potential $V^{+}(x;a_0)$: This, however seems also
impossible, because for $n>0$ the partner potential $V^{+}$ 
obtains new singularities between those of $V^{-}$, corresponding
to the nodes of the $n$-th eigenfunction of $V^{-}$.
Therefore, unless we find some pathological exceptions, it seems
that there is no shape invariance for higher excited states
$n>0$, at least not in the sense of the definition 
(\ref{eq:shapeinvariance}). Perhaps some other functional
relationships yet to be discovered might lead to some other
type of "shape invariance".

\section{The 3-dim spherically symmetric harmonic oscillator}

Let us consider a few examples of susy-m partner potentials, 
first the 3-dim spherically symmetric harmonic oscillator.
(The case of 1-dim harmonic oscillator was calculated and
discussed in (I).)
To prepare some generalities of 3-dim spherically symmetric potentials
$V(r)$ we first write down the Schr\"odinger equation 
$\hat H \psi = E\psi$, namely

\beq
[-\frac{\hbar^2}{2\mu} (\frac{\partial^2}{\partial r^2} + \frac{2}{r}
\frac{\partial}{\partial r}) + \frac{\hat L^2}{2\mu r^2} + V(r)]\psi = E\psi,
\label{eq:radSchrodinger1}
\eeq
where $\mu$ is the mass of the particle and 
$\hat H, \hat L^2$ and $\hat L_z$ are the usual notations for
the Hamilton operator, (square of the) angular momentum, and 
the z-component of the angular
momentum. The quantum numbers of the latter two will be denoted by
$l$ and $m$, so that $l=0,1,2,\dots$ and $m=l, l-1, l-2,\dots, -(l-1),-l$,
and the eigenvalues of $\hat L^2$  are $l(l+1)\hbar^2$.
Due to the spherical symmetry we have thus the separation of variables

\beq
\psi(r,\theta, \varphi) = R(r)Y_{lm}(\theta, \varphi),
\label{eq:separation}
\eeq
where $Y_{lm}$ are the spherical harmonics, so that 

\beq
[-\pr(\frac{d^2}{dr^2} + \frac{2}{r} \frac{d}{dr}) +V(r) +
\frac{l(l+1)\hbar^2}{2\mu r^2}]R = ER,
\label{eq:radSchrodinger2}
\eeq
and after the substitution

\beq
R(r) = \chi(r)/r, \qquad V_l(r) = V(r) + \frac{l(l+1)\hbar^2}{2\mu r^2},
\label{eq:radSchrodinger3}
\eeq
the radial Schr\"odinger equation becomes finally 

\beq
[-\pr \frac{d^2}{dr^2} + V_l(r) ]\chi(r) = E \chi(r),
\label{eq:radSchrodinger4}
\eeq
which is now just a 1-dim problem with the effective potential $V_l(r)$
defined in equation (\ref{eq:radSchrodinger3}), for the physical range
of definition $r\ge0$. From equation (\ref{eq:separation}) follows the
normalization condition

\beq
\int_0^{\infty} dr \chi^2(r) = 1.
\label{eq:normalizationchi}
\eeq

Now we look at the specific case of the 3-dim harmonic oscillator 
defined by 

\beq
V(r) = \half \mu\omega^2 r^2.
\label{eq:harmoscpot}
\eeq
Introducing the constants

\beq
k^2 = \frac{2\mu E}{\hbar^2}, \qquad \lambda^2 = \frac{\mu\omega}{\hbar},
\label{eq:constants1}
\eeq
we rewrite the Schr\"odinger equation (\ref{eq:radSchrodinger4}) as

\beq
\frac{d^2\chi}{dr^2} + [k^2 - \lambda^2r^2 - \frac{l(l+1)}{r^2}] \chi=0,
\label{eq:radharmosc}
\eeq
with the solutions

\beq
\chi_{n}(r) = Cr^{l+1}e^{-\half\lambda^2 r^2} F(-n, l+\frac{3}{2};
\lambda^2r^2),
\label{eq:solharmosc}
\eeq
where $n$ is the radial quantum number $n=0,1,2,\dots$ (in denoting the
eigenfunctions and eigenvalues we shall suppress the quantum number
$l$ for brevity of notation), and the energy eigenvalues read
\beq
E_{n}= (2n+l+\frac{3}{2})\hbar\omega.
\label{eq:energyharmosc}
\eeq
In the above $F(a,b;x)$ is the confluent hypergeometric series,
\footnote{It is also called Kummer's function and denoted by
$M(a,b,z)$ e.g. in Abramowitz and Stegun (1965) p.504.}
and the normalization constant $C$ in (\ref{eq:solharmosc}) 
is equal to (Goldhammer 1963)

\beq
C= \lbrack 
\frac{2^{l+2-n}(2l+2n+1)!!}{\sqrt{\pi}n![(2l+1)!!]^2}\rbrack^{\half}
\lambda^{(l+3/2)}.
\label{eq:normC}
\eeq
The superpotential defined according to (\ref{eq:susyW}) using the
n-th eigenfunction (\ref{eq:solharmosc}) reads

\beq
W_n(r) = -\sqpr \lbrace \frac{l+1}{r} - \lambda^2r \lbrack1+ 
\frac{2n F(-n+1,l+5/2; \lambda^2r^2)}
{(l+3/2)F(-n,l+3/2;\lambda^2r^2)}\rbrack\rbrace
\label{eq:susyWharmosc}
\eeq
so that for the susy-0 (ground state) superpotential we have

\beq
W_0(r) = \sqrt{\frac{\mu}{2}} \omega r - \frac{(l+1)\hbar}{\sqrt{2\mu} r}.
\label{eq:susyWharmosc0}
\eeq
Now we can calculate the partner potential $V_n^{+}$ starting from
the shifted potential

\beq
V_n^{-}(r)=\half \mu\omega^2r^2 + \frac{l(l+1)\hbar^2}{2\mu r^2} 
- (2n+l+3/2)\hbar \omega
\label{eq:harmoscV-}
\eeq
where we see $E_n^{-}=0$, and thus obtain  the partner potential

\beqa
V_n^{+} (r) & = & V_n^{-} (r) + 2\sqpr \frac{dW_n(r)}{dr} =\nonumber\\
            & = & \half\mu\omega^2r^2 + 
\frac{(l+1)(l+2)\hbar^2}{2\mu r^2} - (2n+l+\half)\hbar\omega+I_n, 
\nonumber\\
        I_n & = & \frac{2n\hbar\omega}{(l+3/2)F(0)} 
\lbrace F(1) + \frac{2\mu\omega}{\hbar} r^2 
\lbrack \frac{(1-n)}{(l+5/2)} F(2) + \frac{n}{(l+3/2)} \frac{F^2(1)}{F(0)}
\rbrack\rbrace\nonumber\\
&  &  \nonumber\\
\label{eq:harmoscV+}
\eeqa
where we used the short-hand notation for the confluent hypergeometric
function

\beq
F(i) = F(-n+i, l+\frac{3}{2}+i; \lambda^2r^2), \qquad i=0,1,2.
\label{eq:short-handF}
\eeq
It is readily seen that for $n=0$ we get the shape invariant
case of the two partner potentials

\beqa
V_0^{-}(r,l) & = & \half\mu\omega^2r^2+\frac{l(l+1)\hbar^2}{2\mu r^2} - 
(l+\frac{3}{2})\hbar\omega\nonumber\\
V_0^{+}(r,l) & = & \half\mu\omega^2r^2+\frac{(l+1)(l+2)\hbar^2}{2\mu r^2} - 
(l+\frac{1}{2})\hbar\omega\nonumber\\
V_0^{+}(r,l) & = & V_0^{-}(r,l+1) + 2\hbar\omega\nonumber\\
V_0^{+}(r,l) & = & V_0^{-}(r,a_1) + R(a_1),\nonumber\\
a_0 & = &l, \qquad a_1=l+1,\qquad R(a_1) = 2\hbar\omega.
\label{eq:harmoscshapeinv}
\eeqa
As the last point of this section we give the explicit expressions
for the two lowest excited states ($n=1,2$):

\beqa
V_1^{+}(r) & = & \half\mu\omega^2r^2 + \frac{(l+1)(l+2)\hbar^2}{2\mu r^2}
-(l+\frac{5}{2})\hbar\omega + I_1\nonumber\\
I_1 &= & 2\beta_0\hbar\omega(1-z)^{-1}\lbrack1 + 2\beta_0z(1-z)^{-1}\rbrack
\nonumber\\
\beta_0 & = &(l+\frac{3}{2})^{-1}, \qquad z=\lambda^2r^2 = 
\frac{\mu\omega}{\hbar} r^2 \nonumber\\
\label{eq:harmoscV+1}
\eeqa
and

\beqa
V_2^{+}(r) & = & \half\mu\omega^2r^2 + \frac{(l+1)(l+2)\hbar^2}{2\mu r^2}
-(l+\frac{9}{2})\hbar\omega + I_2\nonumber\\
I_2 & = & 4\beta_0\hbar\omega(1-2\beta_0 z+\beta_0\beta_1z^2)^{-1} \times
\nonumber\\
    & \times & \lbrace 1-\beta_1 z+2z\lbrack-\beta_1 + 2\beta_0
(1-2\beta_0z + \beta_0\beta_1z^2)^{-1}(1-\beta_1z)\rbrack\rbrace
\nonumber\\
\beta_1 & = & (l+\frac{5}{2})^{-1}, \qquad z=\lambda^2r^2 = 
\frac{\mu\omega}{\hbar}r^2 \nonumber\\
\label{eq:harmoscV+2}
\eeqa
The wavefunctions can be calculated using formula (\ref{eq:normalizedpsi}),
where we have

\beq
A = \sqpr \frac{d}{dr} + W_n(r),
\label{eq:harmoscA}
\eeq
and using the eigenfunctions (\ref{eq:solharmosc}) and 
(\ref{eq:susyWharmosc})  we can write down
the expression for the eigenfunctions of the susy-n partner potential,
namely (bearing in mind  $\chi_n^{-} = \chi_{n}$, from 
(\ref{eq:solharmosc})),

\beq
\chi_p^{+} = \frac{1}{\sqrt{E_{n+1+p}^{-}}} A\chi_{n+1+p}^{-},
\label{eq:harmoscwf+}
\eeq
with (using the equation (\ref{eq:energyharmosc}))

\beq
E_{n+1+p}^{-} = E_{n+1+p} - E_{n} = 2(p+1)\hbar\omega,
\qquad p=0,1,2,\dots
\label{eq:harmoscshift}
\eeq

\section{The 3-dim spherically symmetric Kepler problem}

The 3-dim spherically symmetric Kepler potential is

\beq
V(r) = - \frac{e^2}{r}.
\label{eq:Keplerpot}
\eeq
When using this in equations 
(\ref{eq:radSchrodinger1}-\ref{eq:radSchrodinger4})
we find the solutions

\beq
\chi_n(r) = C r^{l+1} e^{-\frac{\lambda}{2}r}F(-n,2l+2;\lambda r),
\label{eq:solKepler}
\eeq
where  $n=0,1,2,\dots$ is the radial quantum number (counting
the number of nodes of $\chi_n(r)$) and the energy eigenvalues
are

\beq
E_n = - \frac{\mu e^4}{2\hbar^2(n+l+1)^2},
\label{eq:energyKepler}
\eeq
with the definitions of the parameters

\beq
\lambda= \frac{2}{(n+l+1)a}, \qquad a=\frac{\hbar^2}{\mu e^2}.
\label{eq:Keplerparameters}
\eeq
The normalization condition is

\beq
\int_0^{\infty}\chi_n^2(r) dr = 1,
\label{eq:normalizationKepler}
\eeq
so that the constant $C$ is equal to

\beq
C= \frac{4}{a^{5/2}N^3(2l+1)!} \sqrt{ \frac{(N+l)!}{(N-l-1)!} },
\qquad N=n+l+1.
\label{eq:CnormalizationKepler}
\eeq
The susy-n superpotential as defined by (\ref{eq:susyW}) reads

\beq
W_n(r) = -\sqpr [\frac{l+1}{r} - \frac{\lambda}{2} - 
\frac{n\lambda F(1)}{(2l+2)F(0)}],
\label{eq:susyWKepler}
\eeq
whose special susy-0 case is

\beq
W_0(r) =\sqrt{\frac{\mu}{2}} \frac{e^2}{(l+1)\hbar}-
\frac{(l+1)\hbar}{\sqrt{2\mu} r}.
\label{eq:susyWKepler0}
\eeq
In the above we used a similar short-hand notation for the
confluent hypergeometric series (function) as in the previous 
section, namely

\beq
F(i) = F(-n+i, 2l+2+i; \lambda r), \qquad i=0,1,2.
\label{eq:shorthandFKepler}
\eeq
Further we calculate the partner potentials

\beqa
V_n^{-}(r) & = & V_l(r) - E_n = -\frac{e^2}{r} + \frac{l(l+1)\hbar^2}{2\mu 
r^2} + \frac{\mu e^4}{2\hbar^2(n+l+1)^2},
\nonumber\\
V_n^{+}(r) & = & -\frac{e^2}{r} + \frac{(l+1)(l+2)\hbar^2}{2\mu 
r^2} + \frac{\mu e^4}{2\hbar^2(n+l+1)^2} + I_n,
\nonumber\\
I_n & = & \frac{n\lambda^2\hbar^2}{\mu(2l+2)F(0)} 
[\frac{(1-n)}{(2l+3)} F(2) + \frac{nF^2(1)}{(2l+2)F(0)}]
\label{eq:KeplerV+}
\eeqa
Now in case of susy-0 partner potnatials we recover the
shape invariance property embodied in the following relationships

\beqa
V_0^{-}(r) & = & -\frac{e^2}{r} + \frac{l(l+1)\hbar^2}{2\mu 
r^2} + \frac{\mu e^4}{2\hbar^2(l+1)^2} \equiv V_0^{-}(r,l),\nonumber\\
V_0^{+}(r) & = & -\frac{e^2}{r} + \frac{(l+1)(l+2)\hbar^2}{2\mu 
r^2} + \frac{\mu e^4}{2\hbar^2(l+1)^2} \equiv V_0^{+}(r,l)\nonumber\\
V_0^{+}(r,l) & = & V_0^{-}(r,l+1) + \frac{\mu e^4}{2\hbar^2}
[\frac{1}{(l+1)^2} - \frac{1}{(l+2)^2}], \nonumber\\
V_0^{+}(r,a_0) & = & V_0^{-}(r,a_1)+R(a_1), \qquad a_o=l,\qquad a_1=l+1,
\nonumber\\
R(a_1) & = & \frac{\mu e^4}{2\hbar^2} [\frac{1}{(l+1)^2} - 
\frac{1}{(l+2)^2}].\nonumber\\
\label{eq:shapeinvKepler}
\eeqa
Finally we calculate the partner potentials for the two
lowest excited states. We find

\beqa
V_1^{+}(r) & = & -\frac{e^2}{r} + \frac{(l+1)(l+2)\hbar^2}{2\mu 
r^2} + \frac{\mu e^4}{2\hbar^2(l+2)^2} + I_1,\nonumber\\
I_1 & = & \frac{\lambda^2\hbar^2}{\mu (2l+2)^2} 
[1 - \frac{\lambda r}{(2l+2)}]^{-2}\nonumber\\
\label{eq:KeplerV+1}
\eeqa
and

\beqa
V_2^{+}(r) & = & -\frac{e^2}{r} + \frac{(l+1)(l+2)\hbar^2}{2\mu 
r^2} + \frac{\mu e^4}{2\hbar^2(l+3)^2} + I_2,\nonumber\\
I_2 & = &  \frac{8\mu e^4\beta_0}{(l+3)^2\hbar^2} 
(1-2\beta0z+\beta_0\beta_1z^2)^{-1}\times\nonumber\\
& \times & [-\beta_1+z\beta_0(1-\beta_1 z)^2
(1-2\beta_0 z+\beta_0\beta_1 z^2)^{-1}],\nonumber\\
\beta_0^{-1} & = & 2l +2,\qquad \beta_1^{-1}=\beta_0^{-1}+1,
\qquad z= \frac{2r}{(l+3)a} = \frac{2\mu e^4r}{(l+3)\hbar^2}.\nonumber\\
\label{eq:KeplerV+2}
\eeqa
Finally we compute the eigenenergies

\beq
E_p^{+} = E_{n+1+p}^{-} = E_{n+1+p} - E_n=
\frac{\mu e^4}{2\hbar^2}[(n+l+1)^{-2} - (n+p+l+2)^{-2}],
\qquad E_0^{+} = E_{n+1}^{-}.
\label{eq:Keplerenergy+}
\eeq
and the eigenfunctions using equations (\ref{operA}), 
(\ref{eq:normalizedpsi}), (\ref{eq:solKepler}), 

\beq
\chi_p^{+}(r) = \frac{1}{\sqrt{E_{n+1+p}^{-}}} A\chi_{n+1+p}^{-}(r),
\qquad \chi_{n+1+p}^{-}(r) = \chi_{n+1+p}(r), \qquad p=0,1,2,\dots
\label{eq:psi+Kepler}
\eeq

\section{The 1-dim Morse potential}

In this section we analyze the supersymmetric aspects of the
Morse potential, defined as

\beq
V(x) = A(e^{-2\alpha x} - 2 e^{-\alpha x}),
\label{eq:Morsepot}
\eeq
which is an important model potential. It has a finite number
of eigenstates. The notation for the constant $A$ here should not
be confused with the (ladder) operator $A$ from section 2.
The Schr\"odinger equation 

\beq
\frac{d^2\psi(x)}{dx^2} + \frac{2\mu}{\hbar^2} (E-V(x))\psi(x) =0,
\label{eq:MorseSchrodinger1}
\eeq
can be rewritten as

\beq
\frac{d^2\psi}{d\xi^2} + \frac{1}{\xi}\frac{d\psi}{d\xi} +
(-\frac{1}{4} + \frac{n+s+\half}{\xi}-\frac{s^2}{\xi^2})\psi=0.
\label{eq:MorseSchrodinger2}
\eeq
where we use the notations

\beq
\xi=\frac{2\sqrt{2\mu A}}{\alpha\hbar}e^{-\alpha x}, \qquad
s=\frac{\sqrt{-\mu E}}{\alpha \hbar},  \qquad n=\frac{\sqrt{2\mu A}}
{\alpha\hbar} -(s+\half).
\label{eq:MorseSchrodinger3}
\eeq
Now we try the Ansatz

\beq
\psi(\xi) = C e^{-\xi/2}\xi^{s} u(\xi), 
\label{eq:MorseSchrodinger4}
\eeq
such that the function $u(\xi)$ must satisfy the simpler equation

\beq
\xi u''+ (2s+1-\xi)u' +nu =0,
\label{eq:MorseSchrodinger5}
\eeq
having the solution in terms of the confluent hypergeometric 
function   

\beq
u(\xi) = F(-n, 2s+1; \xi),
\label{eq:MorseSchrodinger6}
\eeq
so that finally the explicit solution obtains the form

\beq
\psi_n(\xi) = C\xi^{s} e^{-\xi/2}F(-n,2s+1;\xi ),
\label{eq:MorseSchrodingerpsi}
\eeq
with the energy spectrum

\beq
E_{n} = -A[1- \frac{\alpha\hbar}{\sqrt{2\mu A}} (n+\half)]^2,
\qquad n=0,1,2,\dots, < (\frac{\sqrt{2\mu A}}{\alpha\hbar} -\half)
\label{eq:Morseenergy}
\eeq
and the quantum number $n$ runs up to the maximum value.

Now employing the same procedure as before we find the
susy-n superpotential  $W_n(x)$, namely

\beq
W_n(x) = - \sqpr[-\alpha s +\half \alpha\xi +
+ \frac{n\alpha\xi F(1)}{(2s+1) F(0)}],
\label{eq:MorseWn}
\eeq

whose special and well known ground state ($n=0$) value is
\beq
W_0(x) = \sqrt{A} +\frac{\alpha}{2} - \sqrt{A} e^{-\alpha x}.
\label{eq:MorseW0}
\eeq
Again, we use here the short-hand notation for the confluent
hypergeometric function

\beq
F(i) = F(-n+i, 2s+1+i; \xi), \qquad i=0,1,2.
\label{eq:shortF}
\eeq

The starting shifted potential of (\ref{eq:Morsepot}) is

\beq
V_n^{-}(x) = V(x) - E_n = A(e^{-2\alpha x} - 2 e^{-\alpha x}) +
A[1-\frac{\alpha \hbar}{\sqrt{2\mu A}} (n+\half)]^2,
\label{eq:MorseV-}
\eeq
and we get the isospectral partner potential

\beqa
V_n^{+}(x) & = & Ae^{-2\alpha x} - (2 A-\sqrt{\frac{2A}{\mu}}\alpha\hbar)
e^{-\alpha x} + A[1-\frac{\alpha\hbar}{\sqrt{2\mu A}} (n+\half)]^2 + I_n,
\nonumber\\
I_n & = & \frac{n\alpha^2\hbar^2\xi}{(2s+1)\mu F(0)} 
[F(1) + \frac{(1-n)\xi F(2)}{(2s+2)} + \frac{n\xi F^2(1)}{(2s+1)F(0)}].
\nonumber\\
\label{eq:MorseV+}
\eeqa

The shape invariance is recovered for $n=0$, namely the
shifted starting potential is

\beq
V_0^{-}(x) = V(x) - E_0 = A(e^{-2\alpha x} - 2 e^{-\alpha x}) +
A(1-\frac{\alpha \hbar}{2\sqrt{2\mu A}})^2 \equiv V_0^{-}(x;2A),
\label{eq:MorseV-0}
\eeq
and the isospectral partner potential

\beq
V_0^{+}(x) = A e^{-2\alpha x} - (2A-\sqrt{\frac{2A}{\mu}}\alpha\hbar)
 e^{-\alpha x} + A(1-\frac{\alpha \hbar}{2\sqrt{2\mu A}})^2 
\equiv V_0^{+}(x;2A),
\label{eq:MorseV+0}
\eeq
where the notation $V_0^{\pm}(x;2A)$ implies dependence on the coefficient
$2A$ in the Morse potential  (\ref{eq:Morsepot}).
Then we find the shape invariance relationship

\beqa
V_0^{+}(x;a_0) & = & V_0^{-}(x;a_1) + R(a_1),\nonumber\\
a_0 & = & 2A, \qquad a_1=2A-\sqrt{\frac{2A}{\mu}} \alpha\hbar,\nonumber\\
R(a_1) & = & A(1-\frac{\alpha\hbar}{2\sqrt{2A\mu}})^2
\nonumber\\
\label{eq:Morseshapeinv}
\eeqa
The isospectral partner potentials of $V_1^{-}$ and of $V_2^{-}$
are calculated in a straightforward manner:

\beqa
V_1^{+}(x) & = &  Ae^{-2\alpha x} - (2A -\sqrt{\frac{2A}{\mu}}\alpha\hbar) 
e^{-\alpha x} +
A(1-\frac{3\alpha \hbar}{2\sqrt{2\mu A}})^2 + I_1,\nonumber\\
I_1 & = & \frac{\alpha^2\beta_0\hbar\xi}{\mu} (1-\beta_0\xi)^{-1} 
[1+\beta_0\xi (1-\beta_0\xi)^{-1}]\nonumber\\
\label{eq:MorseV1+}
\eeqa
and

\beqa
V_2^{+}(x) & = &  Ae^{-2\alpha x} - (2A -\sqrt{\frac{2A}{\mu}}\alpha\hbar) 
e^{-\alpha x} +
A(1-\frac{5\alpha \hbar}{2\sqrt{2\mu A}})^2 + I_2,\nonumber\\
I_2 & = & 2\alpha^2\beta_0\hbar^2\xi (1-2\beta_0\xi + 
\beta_0\beta_1\xi^2)^{-1} \times \nonumber\\
& \times & [1-\beta_1\xi +2\beta_0\xi 
(1-2\beta_0\xi+\beta_0\beta_1\xi^2)^{-1}(1-\beta_1\xi)^2]\nonumber\\
\label{eq:MorseV+2}
\eeqa
where we use the notation for the parameters $\beta_0,\beta_1$,

\beq
\beta_0= 2s+1, \qquad \beta_1^{-1}=\beta_0^{-1} +1.
\label{eq:Morseparameters}
\eeq
Finally the normalized wavefunctions of the partner potential $V_n^{+}(x)$
are calculated  easily

\beq
\psi_p^{+} = \frac{1}{\sqrt{E_{n+1+p}^{-}}}\hat A\psi_{n+1+p}^{-},
\qquad \psi_{n+1+p}^{-}=\psi_{n+1+p},\qquad E_{n+1+p}^{-}=E_{n+1+p}-E_n,
\label{eq:Morsewf+}
\eeq
where again $p=0,1,2,\dots$, and $\hat A$ here is operator
defined in (\ref{operA}) with the superpotential given in (\ref{eq:MorseWn}).

\section{The 1-dim P\"oschl-Teller type I potential}

The 1-dim P\"oschl-Teller potential is defined as (Fl\"ugge 1971)

\beq
V(x) = \half V_0 [\frac{A(A-1)}{\sin^2\alpha x} +
\frac{B(B-1)}{\cos^2\alpha x}],
\label{eq:PTpot}
\eeq
where 

\beq
V_0 = \frac{\hbar^2\alpha^2}{\mu}, \qquad 0\le \alpha x \le \frac{\pi}{2}.
\label{eq:PTparameters}
\eeq
The solution is well known

\beq
\psi_n(x) = C\sin^A\alpha x \cos^B\alpha x F(-n,A+B+n;\sin^2\alpha x),
\label{eq:solPT}
\eeq
with the energy spectrum

\beq
E_n = \half V_0(A+B+2n)^2,\qquad n=0,1,2,\dots.
\label{eq:energyPT}
\eeq

The susy-n superpotential is

\beqa
W_n(x) & = & -\frac{\alpha \hbar}{\sqrt{2\mu}} [A\cot\alpha x - B\tan \alpha x 
- \frac{2n(A+B+n)\sin \alpha x \cos \alpha x F_n(1)}{(A+\half)F_n(0)]}
\nonumber\\
W_0(x) & = & - \frac{\alpha\hbar}{\sqrt{2\mu}}(A\cot\alpha x -B\tan\alpha x)
\nonumber\\
\label{eq:PTW}
\eeqa
where we employ the short-hand notation for the hypergeometric
series (Abramowitz and Stegun 1965 p.555)

\beq
F_n(i) = F(-n+i, A+B+n+i, A+\half+i; \sin^2\alpha x).
\label{eq:PThypgeoser}
\eeq
The starting shifted potential is

\beq
V_n^{-}(x) = V(x) - E_n = \half V_0 [\frac{A(A-1)}{\sin^2\alpha x} + 
\frac{B(B-1)}{\cos^2\alpha x} - (A+B+2n)^2],
\label{eq:PTV-}
\eeq
and the isospectral partner potential

\beqa
V_n^{+}(x) & = & \half V_0 [\frac{A(A+1)}{\sin^2\alpha x} +
\frac{B(B+1)}{\cos^2\alpha x} - (A+B+2n)^2] + I_n,
\nonumber\\
I_n & = & \frac{2V_0n(A+B+n)}{(A+\half)F_n(0)}
\lbrace F_n(1)(1-2\sin^2\alpha x) + \nonumber\\
    & + &  [\frac{2(-n+1)(A+B+n+1)F_n(2)}{(A+3/2)} +
\frac{2n(A+B+n)F_n^2(1)}{(A+1/2)F_n(0)}] \sin^2\alpha x\cos^2\alpha 
x\rbrace. 
\nonumber\\
\label{eq:PTV+n}
\eeqa
In case $n=0$ we again recover the shape invariance property

\beqa
V_0^{-}(x) & = & \half V_0 [\frac{A(A-1)}{\sin^2\alpha x} + 
\frac{B(B-1)}{\cos^2\alpha x} - (A+B)^2] \equiv V_0^{-}(x,A,B),\nonumber\\
V_0^{+}(x) & = & \half V_0 [\frac{A(A+1)}{\sin^2\alpha x} + 
\frac{B(B+1)}{\cos^2\alpha x} - (A+B)^2] \equiv V_0^{+}(x,A,B),\nonumber\\
V_0^{+} (x,\{a_0\}) & = & V_0^{-}(x,\{a_1\}) + R(\{a_1\}),\nonumber\\
\{a_0\} & = & \{A,B\},\qquad \{a_1\} = \{A+1,B+1\},\qquad 
R(\{a_1\})=V_0(A+B+2).\nonumber\\
\label{eq:PTV+0}
\eeqa

The isospectral partner potentials for the lowest two excited
states $n=1,2$ are 

\beqa
V_1^{+}(x) & = & \half V_0 [\frac{A(A+1)}{\sin^2\alpha x} +
\frac{B(B+1)}{\cos^2\alpha x} - (A+B+2)^2] + I_1,\nonumber\\
I_1 & = & \frac{2V_0(A+B+1)}
{(A+\half)[1 - \frac{(A+B+1)}{(A+\half)}\sin^2\alpha x]} \times\nonumber\\
& \times &
\lbrace 1 - 2\sin^2\alpha x + 
\frac{2(A+B+1)\sin^2\alpha x\cos^2\alpha x}
{(A+\half)[1 - \frac{(A+B+1)}{(A+\half)}\sin^2\alpha x]}  \rbrace
\nonumber\\
\label{eq:PTV+1}
\eeqa
and
\beqa
V_2^{+}(x) & = & \half V_0 [\frac{A(A+1)}{\sin^2\alpha x} +
\frac{B(B+1)}{\cos^2\alpha x} - (A+B+4)^2] + I_2,\nonumber\\
I_2 & = & \frac{4V_0(A+B+2)}{(A+\half)F_2(0)}
\lbrace F_2(1)(1-2\sin^2\alpha x) +\nonumber\\
& + & [-\frac{2(A+B+3)}{(A+3/2)} + \frac{4(A+B+2)F_2^2(1)}{(A+\half)F_2(0)}]
\sin^2\alpha x\cos^2\alpha x\rbrace\nonumber\\
\label{eq:PTV+2}
\eeqa
The procedure to obtain the normalized eigenfunctions of the
isospectral partner potential $V_n^{+}$ is the same as in the
previous sections and in general it is explained in the section 2.

\section{The 1-dim box potential}

The 1-dim box potential is defined as 

\beq
V(x) = 0 \quad {\rm for} -L/2 \le x \le +L/2, \quad {\rm and}\quad V(x)=\infty
\quad {\rm otherwise}.
\label{eq:boxpot}
\eeq
The solutions of the Schr\"odinger equation

\beq
\pr \frac{d^2\psi}{dx^2} + (E-V(x))\psi = 0,
\label{eq:boxSchrodinger1}
\eeq
are

\beqa
\psi_n(x) & = & \sqrt{\frac{2}{L}}\cos \frac{n\pi x}{L},\qquad {\rm  odd}\quad 
n=1,3,,\dots \nonumber\\
\psi_n(x) & = & \sqrt{\frac{2}{L}}\sin \frac{n\pi x}{L},\qquad {\rm even}\quad 
n=2,4 ,\dots \nonumber\\
\label{eq:solbox}
\eeqa
with the energy spectrum

\beq
E_n = \frac{\hbar^2\pi^2}{2\mu L^2} n^2, \qquad n=1,2,\dots.
\label{eq:energybox}
\eeq
The shifted starting potential is
\beq
V_n^{-}(x) = - E_n \quad {\rm for} -L/2 \le x \le +L/2, \quad {\rm and} 
\quad V(x)=\infty \quad {\rm otherwise},
\label{eq:boxV-n}
\eeq
the superpotential is

\beq
W_n(x) = -\sqpr \frac{\psi'_n}{\psi_n},
\label{boxW}
\eeq
and therefore using equation (\ref{eq:VplustoVminus}) we get the
partner potential

\beqa
V_n^{+}(x) & = & -E_n + \frac{\hbar^2}{\mu}\frac{1}{\cos^2 k_nx},
\qquad {\rm odd}, \nonumber\\
V_n^{+}(x) & = & -E_n + \frac{\hbar^2}{\mu}\frac{1}{\sin^2 k_nx},
\qquad {\rm even}, \nonumber\\
k_n  & = & \frac{n\pi}{L}, \qquad E_n=\pr k_n^2, 
\qquad  n=1,2,3,\dots\nonumber\\
\label{eq:boxV+}
\eeqa
Thus the isospectral partner potential to the 1-dim box potential
is a special case of the P\"oschl-Teller type I potential, defined
in (\ref{eq:PTpot}).
The eigenenergies for $V_n^{+}(x)$ are 

\beq
E_{p}^{+} = E_{n+1+p} - E_{n} = \frac{\hbar^2\pi^2}{2\mu L^2}
[(n+1+p)^2-n^2], \qquad p=0,1,2,\dots
\label{eq:energyboxV+}
\eeq
and the eigenfunctions are easily obtained by applying the
formula  (\ref{eq:normalizedpsi}).

\section{Discussion and conclusions}

In this paper we have applied the supersymmetric formalism 
of the (nonrelativistic) quantum mechanics, introduced
and explained in (Robnik 1997), to the higher 
excited states of five different specific exactly solvable
potentials, namely (i) 3-dim (spherically symmetric) harmonic oscillator, 
(ii) 3-dim  (isotropic) Kepler problem, (iii) Morse potential, 
(iv) P\"oschl-Teller type I potential, and 
(v) the 1-dim box potential.  In all cases except in (v) we get
new classes of isospectral partner potentials which thus also fall
into the class of exactly solvable potentials although they
can be and typically are quite complex. The most important
case of the 1-dim harmonic oscillator was treated in detail in
paper (I) (Robnik 1997).
\\\\ 
We have also shown, in section (3), that for higher excited states
there is generally no shape invariance of the usual type,
and it remains to be investigated if there are some other
types of "shape invariance", generated by some other functional
relationships between the starting and partner potentials,
such that they would allow for an exact solution of the problem.
At present we do not know any
specific cases of susy-n shape invariance with $n>0$.
Further calculations for susy-n partner potentials isospectral
to some other well known exactly
solvable potentials, not analyzed in this paper and in (I), 
remain as a future project.

\section*{Acknowledgements}
\par
The financial support by the Ministry of Science
and Technology of the Republic of Slovenia is acknowledged with
thanks.

\newpage

\section*{References} 
\parindent=0. pt
Abramowitz M and Stegun I 1965 {\it Handbook of Mathematical Functions}
(New York: Dover)
\\\\
Barclay D T 1993 {\it Preprint} UICHEP-TH/93-16 Nov 1993
\\\\
Barclay D T, Dutt R, Gangopadhyaya A, Khare A, Pagnamenta A and Sukhatme U
1993 {\it Phys. rev. A} {\bf 48} 2786
\\\\
Barclay D T, Khare A and Sukhatme U 1993 {\it Preprint} UICHEP-TH/93-13
Sept 1993
\\\\
Barclay D T and Maxwell C J 1991 {\it Phys. Lett. A} {\bf 157} 357
\\\\
Courant R and Hilbert D 1968 {\it Methoden der Mathematischen Physik I}
(Berlin: Springer) (in German; translation exists)
\\\\
Dutt R, Khare A and Sukhatme U P 1988 {\it Am. J. Phys.} {\bf 56} 163
(referred to in text as DKS)
\\\\
Fl\"ugge S 1971 {\it Practical Quantum Mechanics I} (Berlin: Springer)
\\\\
Gendenshtein L E 1983 {\it Pisma Zh. Eksp. Teor. Fiz.} {\bf 38} 299
(English translation in {\it JETP Lett.} {\bf 38} (1983) 356)
\\\\
Goldhammer P 1963 {\it Rev. Mod. Phys.} {\bf 35} 40.
\\\\
Infeld L and Hull T E 1951 {\it Rev. Mod. Phys} {\bf 23} 21
\\\\
Inomata A and Junker G 1993 in {\it Lectures in Path Integration: Trieste
1991} eds. H. A. Cerdeira {\em et al} (Singapore: World Scientific)
\\\\
Inomata A and Junker G 1994 {\it Phys. Rev. A} {\bf 50} 3638
\\\\
Inomata A, Junker G and Suparmi A 1993 {\it J. Phys. A: Math. Gen.} {\bf 
26} 2261
\\\\
Junker G 1995 {\it Turk. J. Phys.} {\bf 19} 230
\\\\
L\'evai G 1989 {\it J. Phys. A: Math. Gen.} {\bf 22} 689 
\\\\
Robnik M 1997 {\it J. Phys. A: Math. Gen.} {\bf 30} 1287
\\\\
Robnik M and Salasnich L 1997 {\it J. Phys. A: Math. Gen.} {\bf 30} 1711
\\\\
Schr\"odinger E 1940 {\it Proc. R. Irish Acad.} {\bf A46} 9
\\\\
Schr\"odinger E 1941 {\it Proc. R. Irish Acad.} {\bf A46} 183
\\\\
Witten E 1981 {\it Nucl. Phys. B} {\bf 185} 513
\\\\
\end{document}